\documentclass{article}
\usepackage{spconf,amsmath,graphicx}
\usepackage{booktabs}
\usepackage{graphicx}
\usepackage{xcolor}
\usepackage{soul,color}
\usepackage{url}
\usepackage{float}
\usepackage{enumitem}
\setlist[itemize]{itemsep=2pt, topsep=5pt}

\title{WIKI-EN-ASR-ADAPT: LARGE-SCALE SYNTHETIC DATASET FOR ENGLISH ASR CUSTOMIZATION}
\name{Alexandra Antonova\sthanks{corresponding e-mail: antonova.aleksandra@phystech.edu}}
\address{Moscow Institute of Physics and Technology}

\copyrightnotice{979-8-3503-0689-7/23/\$31.00~\copyright2023 IEEE}
\begin{document}
%
\maketitle
\begin{abstract}
We
present a first large-scale public synthetic dataset for contextual spellchecking customization of automatic speech recognition (ASR) with focus on diverse rare and out-of-vocabulary (OOV) phrases, such as proper names or terms. The proposed approach allows creating millions of realistic examples of corrupted ASR hypotheses and simulate non-trivial biasing lists for the customization task. 
Furthermore, we propose injecting two types of ``hard negatives" to the simulated biasing lists in training examples and describe our procedures to automatically mine them. We report experiments with training an open-source customization model on the proposed dataset and show that the injection of hard negative biasing phrases decreases WER and the number of false alarms.  
\end{abstract}
\begin{keywords}
dataset, personalization, speech recognition, contextual biasing, Wikipedia
\end{keywords}

\begin{figure*}[tp!]
   \centering
   \includegraphics[width=0.85\textwidth]{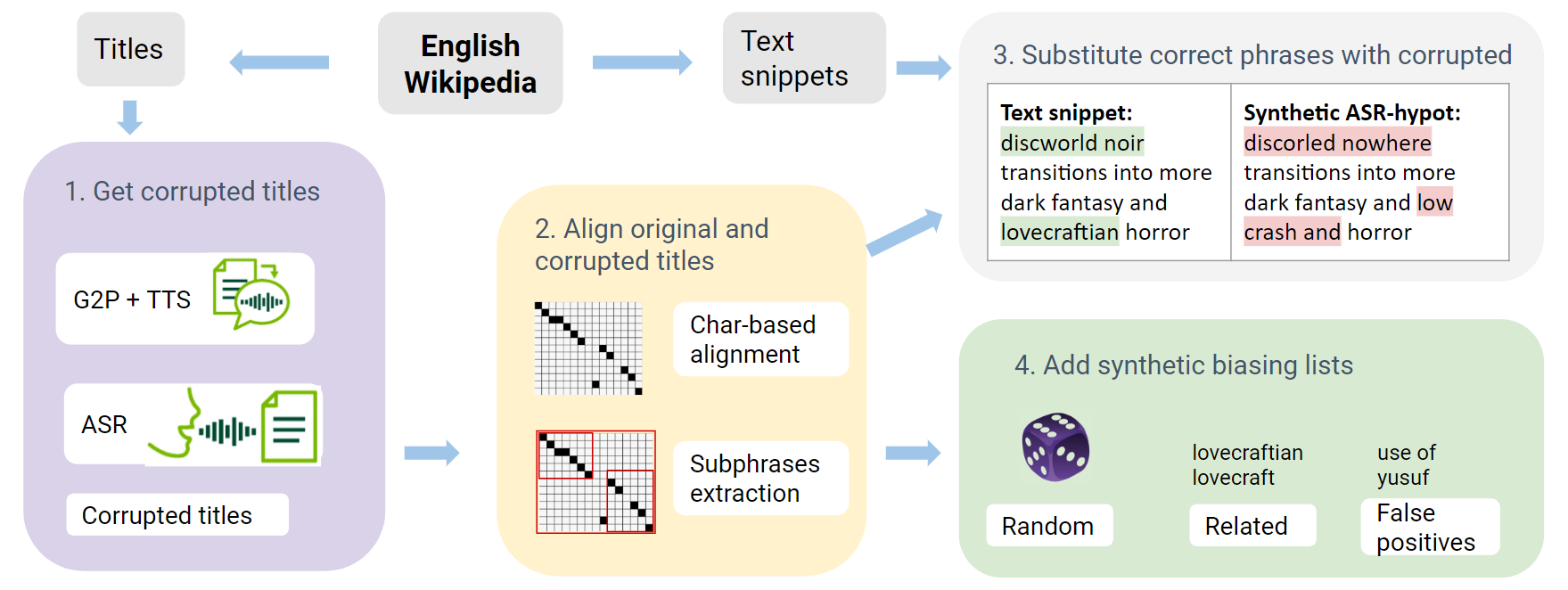}
   \caption {Pipeline for Wiki-En-ASR-Adapt dataset creation using Wikipedia titles and texts.}
\label{fig:pipeline}
\end{figure*}

\section{Introduction}
\label{sec:intro}

Recently, ASR customization problem has been gaining attention from researchers because it is a major issue when using end-to-end ASR models in production systems. While end-to-end ASR models have become popular and typically outperform hybrid ASR models, it is more difficult to boost the correct recognition of user-defined phrases, such as named entities or uncommon terms. The set of all the user's phrases is also referred to as the user's biasing vocabulary, or biasing list.

Different approaches have been proposed. \textit{Shallow fusion} \cite{Zhao2019ShallowFusionEC,gourav2021personalization,fox2022improving} relies on external language models to do on-the-fly rescoring of recognition paths and boost the paths corresponding to biasing phrases. \textit{Deep contextualization} \cite{Pundak2018DeepCE,Jain2020ContextualRF,Chang2021} encodes biasing phrases and integrates them as additional input at training time. Compared to deep contextualization, which requires retraining from scratch, \textit{contextual adapters} \cite{sathyendra2022} are a more lightweight approach that can use a pretrained ASR model with frozen weights and only train a small number of additional parameters. \textit{Text-to-speech (TTS) augmentation} \cite{sim2019personalization,bataev2023textonly} attempts to augment ASR training data with synthesized audio for biasing phrases, but it requires retraining the ASR model, which is impractical except for cases where the model is stored on the user's device. The \textit{contextual spelling correction (CSC)} approach \cite{Wang2022TowardsCS,antonova2023spellmapper,ma2023adapting} deals only with the post-processing of ASR output text. Its task is to correct potentially corrupted ASR output conditioned on the user's biasing list.

The advantage of the CSC approach is that it has the least interaction with the ASR model and can potentially be applied on top of any black-box ASR model, even if it is only accessible for inference via API \cite{ma2023adapting}. The main difficulties are 1) the need to train the correction model from scratch and 2) the lack of training data. Real-world customization data often exist only as in-house datasets, collected by companies providing ASR, e.g. voice assistants, to many users \cite{sathyendra2022,sim2019personalization,Wang2022TowardsCS,Harding2023}. 

Synthetic datasets have proven to be useful for tasks where obtaining natural data is difficult, such as grammar correction \cite{stahlberg-kumar-2021-synthetic} or text normalization \cite{zhang2019neural}. To the best of our knowledge, the only publicly available synthetic dataset for personalization is UserLibri \cite{BreinerRVGMSGCM22}. Though this dataset is based on real data, it is small and is not representative of domains other than fiction. 

The goal of our paper is to construct the first large-scale publicly available synthetic dataset\footnote{dataset available at: \url{https://huggingface.co/datasets/bene-ges/wiki-en-asr-adapt}} for CSC customization. To build such a dataset, we need examples of corrupted ASR transcriptions paired with correct ones and imitations of \textit{user's biasing lists}.

We use Wikipedia titles and their parts as examples of target biasing phrases. This data is large, publicly available and ``universal" since it covers multiple domains frequently needed for customization, such as geographic names, proper names of people from different countries, scientific terminology from different areas, titles of books, movies and so on. We utilize TTS + ASR to obtain examples of ``corrupted" biasing phrases (see Section \ref{ssec:g2p_tts_asr}). To simulate a bigger ``corrupted ASR output" we find occurrences of target phrases in real text snippets from Wikipedia and substitute the correct fragment with its corrupted counterpart. Unlike \cite{Wang2022TowardsCS}, we never insert corrupted phrase at random places, so our training examples always have realistic context around target phrases.

Biasing lists are hard to collect if one does not have access to real user data. It turns out that using \textit{random} biasing lists makes the training task easy, but at inference with real custom vocabulary it may lead to overbiasing - confusion between similar terms and degradation on common words due to false alarms. Some researchers propose injecting \textit{hard negative} examples into biasing lists during training to encourage the neural model to learn more discriminative representations. For example, \cite{Alon2018ContextualSR} proposes adding phonetically similar alternatives, while \cite{bleeker2023approximate} suggests mining hard negative phrases from
the latent space of the context encoder via
approximate nearest neighbour search with a reference query.

In our approach we create two additional sources of \textit{hard negative} biasing phrases (see Section \ref{ssec:negative_examples}): \textit{related} phrases (similar to the correct biasing phrase) and \textit{false positive} phrases (similar to some common words or n-grams). We sample from these sources, in addition to random sampling, when constructing a simulated biasing list for a given training example. 

We aim to make our dataset as universal as possible, providing the possibility to generate final training data with a specified size of biasing lists, ASR-hypothesis length, different sampling strategies, and so on. Our dataset is published in the form of several aggregate tables together with code that allows the creation of training sets with desired properties for different customization task settings. It can serve as a source of diverse training data for models with different architectures. 

Our experiments show that a customization model trained on a 10 million subset of our synthetic dataset improves WER by up to 18\% over the baseline ASR transcription. Hard negative biasing phrases help to reduce the number of false alarms.  

This paper is structured as follows. Section \ref{sec:corpus_preparation} describes corpus preparation from the source data and the tools that we used. In section \ref{sec:biasing_lists}, we propose our approach to hard negative sampling. Section \ref{sec:released_dataset} gives an overview of the structure of the final released dataset. In Section \ref{sec:experiments}, we demonstrate the feasibility of our approach by training an open-source customization model from scratch on a subset of our dataset. Section \ref{sec:conclusion} discusses the results and concludes.


\section{Corpus preparation}
\label{sec:corpus_preparation}
The outline of our approach is shown in Figure \ref{fig:pipeline}. It consists of four steps:
\begin{enumerate}
    \item Feed a large corpus of target biasing phrases to TTS+ASR to get possible corruptions.
    \item Align corrupted and original phrases to extract smaller word n-grams and expand corruption inventory.
    \item Find occurrences in real text snippets to construct training examples.
    \item Add synthetic biasing lists.
\end{enumerate}

\subsection{Running G2P, TTS and ASR to get corrupted (misrecognized) examples}
\label{ssec:g2p_tts_asr}

We follow the approach in \cite{Wang2022TowardsCS,antonova2023spellmapper} and feed a large number of short phrases to TTS and then to an ASR system, to get pairs of input biasing phrase (reference) and the generated ASR-hypotheses, e.g. ``aaron wright" - ``errand right". 

We start with a corpus of 4.5 million raw English Wikipedia titles. We perform some preprocessing, such as removing diacritics, bracketed information, and punctuation. Since there are many out-of-vocabulary words in the titles, we use an external grapheme-to-phoneme (G2P) model (see Table \ref{tab:resources}) to predict the pronunciation of separate words in CMU\footnote{\url{http://www.speech.cs.cmu.edu/cgi-bin/cmudict} } format .
We synthesize audio for each title using TTS with FastPitch mel-spectrogram generator \cite{lancucki2021fastpitch}, which allows us to feed phonemes directly as input, and the HiFi-GAN vocoder\cite{kong2020hifi}. Then, we transcribe the synthesized audio with four different ASR models (Table \ref{tab:resources}) using greedy decoding. 

Table \ref{tab:wer_of_asr_models} shows the metrics of different ASR models after the recognition of synthesized audio for the corpus of Wikipedia titles. We observe that WER of all models is rather high (69-94\%), indicating that this dataset provides many examples of ASR errors. The intersection of predictions is about 20\% between Fast-Conformer-CTC \cite{rekesh2023fast}, Conformer-CTC, Conformer-Transducer \cite{GulatiConformer2020} (Table \ref{tab:resources}), and Whisper models, and about 10\% between each of them and the Quartznet \cite{Kriman2019QuartznetDA} model. Whisper models \cite{radford2022robust} are added only for comparison, we do not use them in the rest of experiments. Whisper models tend to predict overconfidently common phrases instead of what was actually said. On the contrary, Quartznet predictions are often phonetically closer but less consistent with language modeling. The small intersection of predictions of different ASR models means that we can simply increase our corruption dataset by taking predictions of all models on the same input. 

\begin{table}[H]
  \caption{Metrics of different ASR models on synthesized audio for Wikipedia titles.}
  \label{tab:wer_of_asr_models}
  \centering
  \begin{tabular}{rlr}
    \textbf{Model} & \textbf{WER} & \textbf{CER} \\ \midrule
     Conformer-CTC  &  76.17  &  26.7 \\
     FastConformer-CTC  &  76.73  &  27.81  \\
     Conformer-Transducer  &  70.33  &  27.58  \\
     Quartznet  &  94.49  &   34.61  \\
     Whisper-base-en  &  88.18  &  38.86  \\
     Whisper-small-en  &  72.89  &  28.48  \\
     Whisper-medium-en  &  69.35  &  27.0  \\
     Whisper-large  &  69.28  &  27.27  \\ \bottomrule
  \end{tabular}
\end{table}

\subsection{Alignment of original and corrupted titles}
\label{ssec:giza_alignment}
Though there exist many different ways to align a pair of strings, they all lack statistical information. We use stronger corpus-based statistical alignment models provided in the GIZA++ tool \cite{och2003systematic}. Since this tool is designed to align words in a corpus of parallel sentences, we regard our set of original and corrupted titles as ``parallel sentences" and their characters, including spaces, as ``words". 

In \cite{antonova2023spellmapper}, this alignment serves to construct character n-gram mappings but here we use it to extract aligned words and subphrases (word n-grams) to increase the diversity of the corruption inventory. For example, a pair (``ammothea ovatoides", ``amid the overtodes") will give additional paired subphrases (``ammothea", ``amid the") and (``ovatoides", ``overtodes"). 

After this procedure, the initial 4.5 million titles turn into 26 million pairs in the corruption inventory (Table
 \ref{tab:recognition_inventory}). Some of these pairs consist of common frequent words or phrases, which will be filtered later in the pipeline.


\begin{table}[h]
  \caption{Phrase corruption inventory: original (Orig) and recognized (Recog) phrases with counts.}
  \label{tab:recognition_inventory}
  \centering
  \begin{tabular}{llrrllr}
    \textbf{Orig} & \textbf{Recog} & \textbf{Count} & & \textbf{Orig} & \textbf{Recog} & \textbf{Count} \\ \midrule
    congo &  congo &  133  & &  bantu &  band to & 10  \\
    congo &  condo &  9   & &  bantu &  bantu &  9 \\
    congo &  connt go &  1  & &  bantu &  ban to & 7 \\
    congo &  go &     1  & &  bantu &  bant to & 6 \\
    congo &  kango &  1  & &  bantu &  banta &  2 \\
    congo &  come go & 1  & &  bantu &  than too &  1 \\
    congo &  calgo & 1  & &  bantu &  than to &  1  \\
    congo &  kongo &  1  & &  bantu &  bad to &  1  \\ \bottomrule
  \end{tabular}
\end{table}

\subsection{Finding occurrences in real text}
\label{ssec:full_text}
To construct a training example we need not only a correct and corrupted biasing phrase, but also an example of a bigger utterance containing this phrase. To achieve this, we search for the occurrences of reference phrases in real texts.  
We download full texts of all English Wikipedia articles. Those texts are already split into paragraphs, but not into separate sentences. 

Before searching for phrase occurrences, we need to filter out common phrases, to avoid extracting too many paragraphs. We calculate the inverse document frequency (IDF) score for each original and corrupted phrase from our corruption inventory.

\begin{equation}
\label{idf}
    idf(t, D) = log\frac{|D|}{|\{d \in D : t \in d\}|}
\end{equation}

We filter out reference phrases with low IDF scores (e.g. ``in the") or phrases that begin or end with a \textit{word} with low IDF score (too frequent, e.g. ``abraham of", ``in trieste"). 

We use simple text matching in lowercase throughout all Wikipedia articles (not only the article with that particular title). Since the original case is known in the given paragraph, we apply less strict IDF thresholds for uppercase occurrences. We do not use any NER detectors for two reasons: 1) speed and simplicity; 2) some of the biasing phrases that we want to find are not exactly named entities, for example, many scientific terms.

The resulting aggregate table \textbf{Keys2Paragraph} (Table \ref{tab:examples_of_key2paragraph}) has 34 million records consisting of 1) list of reference phrases that occurred in the given paragraph, 2) original paragraph text.

\begin{table}[!tbh]
\caption{Examples from Keys2Paragraph: paragraph text paired with biasing phrases occurring in it.}
\label{tab:examples_of_key2paragraph}

\begin{tabular}{l}
\toprule
\textbf{Keys:} bantu; republic of the congo; congo; mbesa \\
\textbf{Paragraph:} {\color{blue}Mbesa} is a {\color{blue}Bantu} language of the {\color{blue}Democratic} \\
{\color{blue}Republic of the Congo}. \\
\midrule
\textbf{Keys:} mcnab; mcnab bank building; eufaula \\
\textbf{Paragraph:} The {\color{blue}McNab Bank Building} is a historic \\
  building in {\color{blue}Eufaula}, Alabama, U.S.. It was built in the \\
  1850s for John {\color{blue}McNab}, a Scottish-born banker.  \\
 \bottomrule

\end{tabular}
\end{table}

\subsection{Simulation of ASR-hypothesis and positive biasing phrases}

Based on Keys2Paragraph, ASR-hypotheses can be easily created by cutting text snippets of arbitrary length and substituting reference phrases with some of their counterparts from the corruption inventory (see Table \ref{tab:recognition_inventory}). Note that the described substitution procedure allows us to construct highly realistic examples of ASR-hypotheses with misrecognized biasing phrases (Table \ref{tab:examples_of_asr_hypotheses}), because 
\begin{enumerate}
\item The substituted phrases are taken from titles or their parts, not just random word n-grams, and are likely to represent some named entity or term. 
\item The corrupted variants are produced by real ASR systems.
\item The corrupted variant is used exactly in place of the original phrase and not at a random place in the sentence like in \cite{Wang2022TowardsCS}.
\end{enumerate}

\begin{table}[!tbh]
\caption{Examples of synthetic ASR-hypotheses: text snippets with correct biasing phrases substituted by misrecognized.}
\label{tab:examples_of_asr_hypotheses}

\begin{tabular}{l}
\toprule
\textbf{Original:} {\color{blue}discworld noir} transitions into more dark \\
fantasy and {\color{blue}lovecraftian} horror    \\
\textbf{Corrupted:} {\color{red}discorled nowhere} transitions into more dark \\
fantasy and {\color{red}low crash and} horror    \\
\midrule
\textbf{Original:} of a dog with {\color{blue}endocarditis} due to \\
{\color{blue}bartonella rochalimae} was published in \\
\textbf{Corrupted:} of a dog with {\color{red}andocrditis} due to \\
{\color{red}bartonel rokalima} was published in    \\
\bottomrule
\end{tabular}
\end{table}

The original fragments that were substituted with a corrupted variant serve as the \textit{positive} biasing phrases for this particular example. There is a question of whether one should consider ``self-replacement" as valid biasing example. In our experiments, we allow for such self-replacements so that model learns to detect them along with corrupted variants. This can be controlled at the stage of the generation of the final dataset.



\subsection{Fast dictionary-based text normalization}
Since many ASR models output normalized text, to imitate ASR-hypotheses, we may need to normalize the paragraph texts before cutting snippets. It can be time-consuming to process such a large amount of text by neural or complex rule-based text normalization systems. We developed a fast dictionary-based text normalization. It remembers all 4.3 million normalized/unnormalized phrase equivalents from English Google Text Normalization Dataset \cite{zhang2019neural} and just replaces matching word n-grams in the paragraph text with their most frequent normalization equivalent. The procedure always tries to match the longest possible n-grams first. In rare cases, some text fragments may remain unnormalized (e.g. containing a long unknown number). Such fragments are skipped during snippet selection. Skipping is acceptable because we have abundant data.
 
\section{Simulation of biasing lists}
\label{sec:biasing_lists}
The procedure described in Section \ref{sec:corpus_preparation} allows us to obtain a realistic text snippet (ASR-hypothesis) with corrupted fragments and one or more \textit{positive} biasing phrases. However, we have to somehow add biasing lists to the training data in order to make it closer to real data at inference. This is not easy because there exist no public datasets with biasing lists, and there is no known way to extract them from any available data.
That is why researchers put effort into simulating biasing lists \cite{Alon2018ContextualSR, bleeker2023approximate}. 

\subsection{Types of negative examples}
\label{ssec:negative_examples}
The customization task is substantially different from other language modeling tasks because it focuses on cases that are hard or impossible to learn even from large amounts of text:
\begin{enumerate}
\item Rare or out-of-vocabulary phrases (e.g. ``aaadonta fuscozonata"). 
\item Undistinguishable spelling variations (e.g. ``nathalie", ``nataly").
\end{enumerate}
This means that the model should learn to \textit{phonetically} compare biasing phrases to the text of ASR-hypothesis. At the same time, it can rely on information about the meaning of the surrounding words. For example, it can notice meaning inconsistencies, which may indicate that this is a corrupted fragment.

We consider the following types of useful negative biasing phrases:
\begin{enumerate}
\item \textit{Random} - Just random biasing phrases from a big pool. 
\item \textit{Related} - Phrases that are somewhat similar to a positive biasing phrase, but distinguishable from it (e.g. ``lovecraftian"/``lovecraft", ``mizukawa"/``kurizuka", ``boulter"/``ana boulter", ``anaudia"/``antony naudi").
\item \textit{False positives} - Phrases that are accidentally similar to some valid fragment in the ASR-hypothesis, that need not be replaced (e.g. ``nuts and"/``knutsen", ``use of"/``yusuf", ``helmet"/``hellmut", ``runs"/``rhoannes", ``application"/``plication", ``difference"/``deference").
\end{enumerate}

Note that spelling variations (e.g. ``nathalie", ``nataly") are useful as positive biasing phrases, but \textit{not} useful as negatives. This would correspond to a situation in which a user's vocabulary has both ``nathalie" and ``nataly", and they would be indistinguishable to the model because they are close both phonetically and semantically.


\subsection{Finding ``related" phrases}
\label{ssec:related_phrases}

The purpose of the \textit{related} class of negative examples is to serve as distractors, making the recognition task harder for the model and teaching it to capture subtle differences to better discriminate between similar phrases. In real settings, a custom vocabulary can often contain related terms, such as ``cardiac" / ``cardial" / ``cardiology",  ``tubercular" / ``tuberculous" / ``tuberculosis", and we don't want the model to confuse them with one another.

\subsection{Finding false positives}
\label{ssec:false_positives}
The purpose of the \textit{false positive} class of negative examples is to teach the model to distinguish cases when some fragment in the ASR-hypothesis is accidentally phonetically similar to some biasing phrase, but is relevant to its surrounding context and need not be replaced. For example, in the text snippet ``of filo pastry filled with chopped nuts and soaked in honey", the fragment ``nuts and" is perfectly relevant and need not be replaced with biasing phrase ``knutsen". On the contrary, in the text snippet ``hartman lived with nuts and and her two sisters" the fragment ``nuts and" is irrelevant which is a hint for the model that it's a corrupted biasing phrase ``knutsen", provided it is in the biasing list. Note that despite the fact that the biasing phrase itself can be completely unseen, models can learn such differences by looking at how well the corrupted fragment fits into the surrounding context. The only problem is that it's hard to sample a sufficient number of false positive pairs like ``knutsen" and ``nuts and" because it's combinatorially difficult to compare the set of possible biasing phrases to the set of possible text n-grams. 

Our solution for mining false positive pairs is as follows: we take all phrases from our corruption inventory and find occurrences of \textit{corrupted} variants in Wikipedia texts. Since the text is real, we consider the corrupted fragment to be relevant - and therefore we can use the original phrase as a false positive example for the given text snippet.

Unfortunalely, this method cannot generate an arbitrarily large number of false positive examples, because it is restricted by the phrases from our corruption inventory, many of which are not realistic n-grams and do not occur in real texts.
Our final false positive phrases inventory contains 449 thousand pairs for 66 thousand common n-grams.

\begin{table}[t]
\caption{Links to resources used in this work.}
\label{tab:resources}
\centering
\resizebox{\columnwidth}{!}{%
\begin{tabular}{ll}
\textbf{Resource} & \textbf{Reference} \\
\midrule
G2P & \textit{https://huggingface.co/bene-ges/en\_g2p\_cmu\_bert\_large} \\
\midrule
& \textit{https://catalog.ngc.nvidia.com/models}:                 \\
Conformer-CTC & stt\_en\_conformer\_ctc\_large \\
Conformer-RNNT & stt\_en\_conformer\_transducer\_large \\
FastConformer-CTC & stt\_en\_fastconformer\_ctc\_large \\
Quartznet15x5 & stt\_en\_quartznet15x5 \\
Mel-spectrogram & tts\_en\_fastpitch \\
Vocoder & tts\_hifigan \\
\midrule
& \textit{https://github.com}:                 \\
100k medical terms & glutanimate/wordlist-medicalterms-en \\
24k multi-word medical terms & McGill-NLP/medal/master/toy\_data/valid\_adam.txt \\
\bottomrule
\end{tabular}%
}
\end{table}

\section{Description of the released dataset}
\label{sec:released_dataset}
Wiki-En-ASR-Adapt dataset is published as several aggregate tables together with code that allows the creation of training sets with desired properties for different customization task settings. The resulting resource contains:
\begin{itemize}
\item \textbf{Keys2Paragraph} - 4.3 million unique words/phrases occurring in 33.8 million paragraphs. 
\item \textbf{Keys2Corruptions} - 24.7 million pairs in the corrupted phrase inventory, as recognized by four different ASR models.
\item \textbf{Keys2Related} - 62.7 million pairs in the related phrase inventory.
\item \textbf{FalsePositives} - 449 thousand pairs in the false positive phrase inventory.
\item Scripts to generate training examples with the required properties and sampling strategy.
\item Wikipedia titles with character-level alignment, done with GIZA++.
\end{itemize}

\section{Experiments}
\label{sec:experiments}

Most of the existing customization models are in-house systems of commercial companies. We use an open-source customization model\cite{antonova2023spellmapper} from the NeMo toolkit \cite{kuchaiev2019nemo}. We generate training data in the required format using our Wiki-En-ASR-Adapt dataset and verify that the model can learn from scratch on it.

\subsection{SpellMapper model architecture}
\label{ssec:spellmapper}
SpellMapper is a non-autoregressive model based on BERT architecture that works at char-level and \textit{tags} characters as belonging to one of ten given candidate biasing phrases or none. The given ten biasing phrases are preselected from a larger biasing vocabulary by a retrieval algorithm that finds candidates that best match the input ASR-hypothesis based on n-gram mappings statistics. 

\subsection{Training and evaluation datasets}
\label{ssec:datasets}
We use Wiki-En-ASR-Adapt to generate two training datasets  for the SpellMapper model. Both datasets consist of ten million training examples, and in 50\% of examples there is at least one correct candidate. Biasing lists always contain ten phrases. In the first dataset, all phrases in the biasing list (except for the correct candidates) are sampled randomly from a big pool of phrases. In the second dataset, we inject hard negative biasing phrases in the following way:
\begin{itemize}
\item 1-3 incorrect candidates are sampled from the false positives inventory if available.
\item If the example contains correct candidate(s), at most 3 incorrect candidates are sampled from the related phrases inventory.
\item The rest are randomly sampled to reach the total number of ten candidates.
\end{itemize}

For the final testing, we use three corpora from different domains. Spoken Wikipedia \cite{Baumann2019TheSW} and SPGISpeech \cite{oneillSPGISpeech} are processed similarly to \cite{antonova2023spellmapper}: audio is real, and biasing vocabularies are simulated using rare words and phrases from the reference texts. To test with a more realistic biasing vocabulary, we prepare a medical dataset (Med) in the following way: we take a set of 5.5k medical abstracts from PubMed\footnote{\url{https://www.ncbi.nlm.nih.gov/pmc/}} resource and intersect their texts with a list of medical terms consisting of 100k single-word terms and 24k multi-word terms (see resources in Table \ref{tab:resources}), obtaining a biasing vocabulary of 5.1k phrases. Then we feed the medical abstracts to TTS (same as in Section \ref{ssec:g2p_tts_asr}) to get synthetic audio. 

Baseline transcriptions for all three datasets are produced by the Conformer-CTC model. Apart from Word Error Rate (WER), we measure recall and precision w.r.t. the target biasing phrases and the percentage of transcriptions changed by post-correction. We compare the results of the customization model trained on datasets with and without hard negative biasing phrases.

    

\subsection{Results}
\label{ssec:results}

The results are presented in Table \ref{tab:results}. They can be summarized as follows:
\begin{itemize}
\item The proposed dataset allows to train a working customization model from scratch.
\item Model trained with random biasing lists tends to make more replacements at inference but has low precision, diminishing gain in WER. 
\item Injection of hard negative biasing phrases decreases WER and affects the precision/recall metrics making the model produce fewer false alarms.
\end{itemize}

\begin{table}[H]
  \caption{Comparison of customization models trained with random/hard biasing lists, using test sets from different domains: medical (Med), business (SPGI), Wikipedia (SWC).}
  \label{tab:results}
  \centering
  \begin{tabular}{llllr}
    \multicolumn{1}{c}{\begin{tabular}[c]{@{}c@{}}\textbf{Training}\\ \textbf{set}\end{tabular}} & \textbf{Metric} & \textbf{Med} & \textbf{SPGI} & \textbf{SWC} \\ \midrule
    Random & WER &  11.46  & 5.64  & 5.54 \\
    biasing & Recall &  58.8  &  61.3  &  68.1 \\
    lists & Precision &  53.2  &  65.1  & 73.0 \\
    & Changed sent &  65.8  &   10.0  & 30.3 \\
     \midrule
    Hard & WER  &  10.77  &  5.59 &  5.39 \\
    biasing & Recall &  50.2  &  46.1  & 64.6 \\
    lists & Precision &  63.2  &  84.6  & 81.4 \\
    & Changed sent &  52.8  &   6.0  &  25.7 \\
     \midrule
    Baseline ASR & WER & 12.72 & 5.88 &  6.6 \\ 
     \bottomrule
  \end{tabular}
\end{table}




\section{CONCLUSION}
\label{sec:conclusion}
In this paper we introduced Wiki-En-ASR-Adapt, the first large-scale dataset for English ASR customization with focus on rare and OOV biasing phrases. Despite being synthetic, our training examples are realistic because they are based on real text snippets. The large corpus size, including multiple domains covered by Wikipedia, and the use of real errors of different ASR models make it representative for many potential customization applications. 

We also presented a solution to ``hard negative" problem. We supply two sources of ``hard negatives", namely, \textit{related phrases} and \textit{false positives}, and provide an option to inject them in desired proportions to the simulated biasing lists in training examples. 

We demonstrated the feasibility of the proposed approach by training an open-source customization model on the generated dataset and showed the positive effect of hard negatives.

We hope that the appearance of a new large-scale public dataset for ASR customization will stimulate academic research and development of new customization models. While our dataset is best suited to train text-only CSC customization models, like \cite{Wang2022TowardsCS} or \cite{antonova2023spellmapper}, it can be adapted to training customization models that use acoustics if one synthesizes audio for selected sentences.

Current restrictions that have yet to be overcome:
\begin{itemize}
\item Since our corruption inventory originates from Wikipedia titles, which are mostly noun groups in singular grammatical number, other classes, such as verbs, -ing, possessive, and plural forms, are underrepresented. 
\item False positive inventory should be expanded to cover more common word n-grams.
\end{itemize}

All code to reproduce our experiments will be released together with the dataset.

\bibliographystyle{IEEEbib}
\bibliography{strings,refs}

\begin{thebibliography}{10}

\bibitem{Zhao2019ShallowFusionEC}
Ding Zhao, Tara~N. Sainath, David Rybach, Pat Rondon, Deepti Bhatia, Bo~Li, and
  Ruoming Pang,
\newblock ``Shallow-fusion end-to-end contextual biasing,''
\newblock in {\em Interspeech}, 2019.

\bibitem{gourav2021personalization}
Aditya Gourav, Linda Liu, Ankur Gandhe, Yile Gu, Guitang Lan, Xiangyang Huang,
  Shashank Kalmane, Gautam Tiwari, Denis Filimonov, Ariya Rastrow, et~al.,
\newblock ``Personalization strategies for end-to-end speech recognition
  systems,''
\newblock in {\em ICASSP}, 2021.

\bibitem{fox2022improving}
Jennifer~Drexler Fox and Natalie Delworth,
\newblock ``Improving contextual recognition of rare words with an alternate
  spelling prediction model,''
\newblock in {\em Interspeech}, 2022.

\bibitem{Pundak2018DeepCE}
Golan Pundak, Tara~N. Sainath, Rohit Prabhavalkar, Anjuli Kannan, and Ding
  Zhao,
\newblock ``Deep context: End-to-end contextual speech recognition,''
\newblock {\em 2018 IEEE Spoken Language Technology Workshop (SLT)}, pp.
  418--425, 2018.

\bibitem{Jain2020ContextualRF}
Mahaveer Jain, Gil Keren, Jay Mahadeokar, Geoffrey Zweig, Florian Metze, and
  Yatharth Saraf,
\newblock ``Contextual {RNN-T} for open domain {ASR},''
\newblock in {\em Interspeech}, 2020.

\bibitem{Chang2021}
Feng-Ju~(Claire) Chang, Jing Liu, Martin Radfar, Athanasios Mouchtaris,
  Maurizio Omologo, Ariya Rastrow, and Siegfried Kunzmann,
\newblock ``Context-aware transformer transducer for speech recognition,''
\newblock in {\em ASRU}, 2021.

\bibitem{sathyendra2022}
Kanthashree~Mysore Sathyendra, Thejaswi Muniyappa, Feng-Ju~(Claire) Chang, Jing
  Liu, Jinru Su, Grant Strimel, Athanasios Mouchtaris, and Siegfried Kunzmann,
\newblock ``Contextual adapters for personalized speech recognition in neural
  transducers,''
\newblock in {\em ICASSP 2022}, 2022.

\bibitem{sim2019personalization}
Khe~Chai Sim, Fran{\c{c}}oise Beaufays, Arnaud Benard, Dhruv Guliani, Andreas
  Kabel, Nikhil Khare, Tamar Lucassen, Petr Zadrazil, Harry Zhang, Leif
  Johnson, et~al.,
\newblock ``Personalization of end-to-end speech recognition on mobile devices
  for named entities,''
\newblock in {\em Automatic Speech Recognition and Understanding Workshop
  (ASRU)}, 2019.

\bibitem{bataev2023textonly}
Vladimir Bataev, Roman Korostik, Evgeny Shabalin, Vitaly Lavrukhin, and Boris
  Ginsburg,
\newblock ``Text-only domain adaptation for end-to-end asr using integrated
  text-to-mel-spectrogram generator,''
\newblock in {\em Interspeech}, 2023.

\bibitem{Wang2022TowardsCS}
Xiaoqiang Wang, Yanqing Liu, Jinyu Li, Veljko Miljanic, Sheng Zhao, and Hosam
  Khalil,
\newblock ``Towards contextual spelling correction for customization of
  end-to-end speech recognition systems,''
\newblock {\em IEEE/ACM Transactions on Audio, Speech, and Language
  Processing}, vol. 30, pp. 3089--3097, 2022.

\bibitem{antonova2023spellmapper}
Alexandra Antonova, Evelina Bakhturina, and Boris Ginsburg,
\newblock ``Spellmapper: A non-autoregressive neural spellchecker for asr
  customization with candidate retrieval based on n-gram mappings,''
\newblock in {\em Interspeech}, 2023.

\bibitem{ma2023adapting}
Rao Ma, Mengjie Qian, Mark J.~F. Gales, and Kate~M. Knill,
\newblock ``Adapting an unadaptable asr system,''
\newblock in {\em Interspeech}, 2023.

\bibitem{Harding2023}
Philip Harding, Sibo Tong, and Simon Wiesler,
\newblock ``Selective biasing with trie-based contextual adapters for
  personalised speech recognition using neural transducers,''
\newblock in {\em Interspeech}, 2023.

\bibitem{stahlberg-kumar-2021-synthetic}
Felix Stahlberg and Shankar Kumar,
\newblock ``Synthetic data generation for grammatical error correction with
  tagged corruption models,''
\newblock in {\em Proceedings of the 16th Workshop on Innovative Use of NLP for
  Building Educational Applications}, Online, Apr. 2021, pp. 37--47,
  Association for Computational Linguistics.

\bibitem{zhang2019neural}
Hao Zhang, Richard Sproat, Axel~H Ng, Felix Stahlberg, Xiaochang Peng, Kyle
  Gorman, and Brian Roark,
\newblock ``Neural models of text normalization for speech applications,''
\newblock {\em Computational Linguistics}, vol. 45, no. 2, pp. 293--337, 2019.

\bibitem{BreinerRVGMSGCM22}
Theresa Breiner, Swaroop Ramaswamy, Ehsan Variani, Shefali Garg, Rajiv Mathews,
  Khe~Chai Sim, Kilol Gupta, Mingqing Chen, and Lara McConnaughey,
\newblock ``{UserLibri: A dataset for ASR personalization using only text},''
\newblock in {\em Interspeech}, 2022.

\bibitem{Alon2018ContextualSR}
Uri Alon, Golan Pundak, and Tara~N. Sainath,
\newblock ``Contextual speech recognition with difficult negative training
  examples,''
\newblock {\em ICASSP 2019 - 2019 IEEE International Conference on Acoustics,
  Speech and Signal Processing (ICASSP)}, pp. 6440--6444, 2018.

\bibitem{bleeker2023approximate}
Maurits Bleeker, Pawel Swietojanski, Stefan Braun, and Xiaodan Zhuang,
\newblock ``Approximate nearest neighbour phrase mining for contextual speech
  recognition,'' 2023.

\bibitem{lancucki2021fastpitch}
Adrian {\L}a{\'n}cucki,
\newblock ``{FastPitch}: Parallel text-to-speech with pitch prediction,''
\newblock in {\em ICASSP}, 2021.

\bibitem{kong2020hifi}
Jungil Kong, Jaehyeon Kim, and Jaekyoung Bae,
\newblock ``{Hifi-GAN}: Generative adversarial networks for efficient and high
  fidelity speech synthesis,''
\newblock in {\em NeurIPS}, 2020.

\bibitem{rekesh2023fast}
Dima Rekesh, Samuel Kriman, Somshubra Majumdar, Vahid Noroozi, He~Huang,
  Oleksii Hrinchuk, Ankur Kumar, and Boris Ginsburg,
\newblock ``Fast conformer with linearly scalable attention for efficient
  speech recognition,'' 2023.

\bibitem{GulatiConformer2020}
Anmol Gulati, James Qin, Chung{-}Cheng Chiu, Niki Parmar, Yu~Zhang, Jiahui Yu,
  Wei Han, Shibo Wang, Zhengdong Zhang, Yonghui Wu, and Ruoming Pang,
\newblock ``Conformer: Convolution-augmented transformer for speech
  recognition,''
\newblock in {\em Interspeech}. 2020, {ISCA}.

\bibitem{Kriman2019QuartznetDA}
Samuel Kriman, Stanislav Beliaev, Boris Ginsburg, Jocelyn Huang, Oleksii
  Kuchaiev, Vitaly Lavrukhin, Ryan Leary, Jason Li, and Yang Zhang,
\newblock ``Quartznet: Deep automatic speech recognition with 1d time-channel
  separable convolutions,''
\newblock {\em ICASSP 2020 - 2020 IEEE International Conference on Acoustics,
  Speech and Signal Processing (ICASSP)}, pp. 6124--6128, 2019.

\bibitem{radford2022robust}
Alec Radford, Jong~Wook Kim, Tao Xu, Greg Brockman, Christine McLeavey, and
  Ilya Sutskever,
\newblock ``Robust speech recognition via large-scale weak supervision,''
\newblock in {\em International Conference on Machine Learning}, 2022.

\bibitem{och2003systematic}
Franz~Josef Och and Hermann Ney,
\newblock ``A systematic comparison of various statistical alignment models,''
\newblock {\em Computational linguistics}, vol. 29, no. 1, pp. 19--51, 2003.

\bibitem{kuchaiev2019nemo}
O.~Kuchaiev, J.~Li, H.~Nguyen, O.~Hrinchuk, R.~Leary, B.~Ginsburg, S.~Kriman,
  S.~Beliaev, V.~Lavrukhin, J.~Cook, et~al.,
\newblock ``{NeMo}: a toolkit for building {AI} applications using neural
  modules,''
\newblock in {\em {Systems for ML Workshop, NeurIPS}}, 2019.

\bibitem{Baumann2019TheSW}
Timo Baumann, Arne K{\"o}hn, and Felix Hennig,
\newblock ``The {Spoken Wikipedia} corpus collection: Harvesting, alignment and
  an application to hyperlistening,''
\newblock {\em Language Resources and Evaluation}, vol. 53, pp. 303--329, 2019.

\bibitem{oneillSPGISpeech}
Patrick~K. O'Neill, Vitaly Lavrukhin, Somshubra Majumdar, Vahid Noroozi, Yuekai
  Zhang, Oleksii Kuchaiev, Jagadeesh Balam, Yuliya Dovzhenko, Keenan Freyberg,
  Michael~D. Shulman, Boris Ginsburg, Shinji Watanabe, and Georg Kucsko,
\newblock ``{SPGISpeech}: 5, 000 hours of transcribed financial audio for fully
  formatted end-to-end speech recognition,''
\newblock in {\em Interspeech}, 2021.

\end{thebibliography}

\end{document}